# Identifying Topological Phase Transitions in Experiments Using Manifold Learning

**Eran Lustig\*, Or Yair\*, Ronen Talmon, Mordechai Segev**

Technion – Israel Institute of Technology, Haifa 32000, Israel
\*Equal contribution

We demonstrate the identification and classification of topological phase transitions from experimental data using Diffusion Maps: a nonlocal unsupervised machine learning method. We analyze experimental data from an optical system undergoing a topological phase transition and demonstrate the ability of this approach to identify topological phase transitions even when the data originates from a small part of the system, and does not even include edge states.

Topological phases are currently at the heart of modern research in multiple areas in physics due to their fundamental nature and numerous potential applications. These phases are characterized by unique properties such as robust unidirectional edge-states, exhibiting transport that is immune to scattering and disorder, and quantized conductance [1–3]. These distinct features were observed in many experiments and on several different platforms [4–10]. However, numerous topological phases were predicted but thus far never observed in any experiment [11,12], and new topological phases are still being proposed theoretically [12], with the hope to be observed in the future. Experimentally, topological phases are typically identified by studying the features of the edge transport in the underlying medium, through conductance experiment (in solid-state), or by imaging the transport of light waves, acoustic waves, etc. This creates major challenges because in many cases the system does not have a clear edge (e.g., cold atoms in magneto-optic traps), or because only bulk states are accessible in experiments (for various reasons, e.g., only part of the spectrum is accessible, etc.). The challenge in observing topological phase transitions stems from their nonlocal nature, that is: the non-trivial topology is a property of the entire system, and not of a local part of it. As a result, topological phase transitions can often be revealed only by specific excitations that require both a high degree of control over the system and a firm understanding of it [8,9].

Furthermore, modern experimental systems can produce large amounts of data with many degrees of freedom that are impossible to exploit without automation. These and other challenges exemplify the need for machine learning algorithms to identify and characterize topological phase transitions, based on partial (incomplete) experimental data. In principle, neural networks can be trained on known phase transitions to detect an unknown phase transition in the Hamiltonian [13] or the entanglement spectrum [14], as was demonstrated on experimental data [15], and on numerous cases based on data from simulation [13,14,16–20]. However, these methods rely on prior knowledge on the physical system itself, which is fundamentally problematic when trying to reveal new and unknown phase transitions and new phases of matter. This is manifested in the training process, which is based on known phase transitions. To address new phase transitions, where one cannot rely on previous knowledge acquired on other (known) systems, it was proposed to use unsupervised learning algorithms, i.e., learning from the data directly, without relying on any labeling. In machine learning, unsupervised learning was demonstrated to cluster data based on certain features of the systems [21,22]. Without prior knowledge, these approaches generally have limited success: on the one hand, unsupervised learning does not have a bias towards known phenomena, yet on the other hand, it does not produce high-quality distinction [23]. To overcome these challenges, it was recently suggested [24,25] to use a manifold-learning methodology called Diffusion Maps [26]. In a very recent theory paper [25], diffusion maps were shown numerically to retrieve the Thouless phase transition based on simulated measurements. However, it is still unclear whether this approach can handle true experimental data, where the noise is not well characterized and is often not homogeneous throughout the system, and most importantly – in many cases the edge states or parts of the spectrum are inaccessible to experiments.

Here, we use Diffusion Maps to identify the topological phase transition of a photonic system from true experimental data, without any prior assumptions on the system. We show that Diffusion Maps can characterize the phase transition of a topological system, based on experimental excitations that do not include edge-states. We also show that identifiction based on random excitations or excitations that are limited to a small part of the system's spectrum is attainable. Finally, because this system is engineered to demonstrate phase transitions upon varying known parameters, we compare the performance of the Diffusion Maps methodology to the true phase transitions and conclude that indeed this technique can identify topological phase transition from measurements, based on the data itself without any prior knowledge.

Let us briefly introduce Diffusion Maps, which is a kernel-based method for manifold learning. Generally, given a set of $N$ consecutive snapshots of the evolution of a dynamical system $\{x_i \in \mathbb{R}^D\}_{i=1}^N$, where $i$ represents different consecutive sampling times of the observed data $x$ and $D$ is the dimension of each snapshot, Diffusion Maps seeks a natural low-dimensional representation of the observations, which should be isomorphic to the latent underlying intrinsic state of the system. In this work, we demonstrate that the low-dimensional representation obtained via Diffusion Maps facilitates the identification of topological phase transitions from experimental data, with high precision. Formally, given the set $\{x_i\}$, we construct the kernel $K[i,j] = \exp\left(-\frac{\|x_i - x_j\|^2}{2\sigma^2}\right)$ for all $i, j \in \{1, 2, \ldots, N\}$, where $\|\cdot\|$ is a suitable norm operator. Here, $\sigma$ is the Gaussian kernel scale which can be optimized for each model, but we set it consistently to be 5% of the median of all pairwise differences between the wavefunctions, i.e., $\sigma = 0.05 \cdot Median\left(\{\|x_i - x_j\|\}_{i,j}\right)$. A common practice is to use the Euclidean norm, but other norms can be used as well. We then normalize $K$ to be row stochastic according to $A[i,j] = \frac{K[i,j]}{\sum_{s=1}^N K[i,s]}$. Next, we apply eigendecomposition such that $A = \Phi \Lambda \Phi^{-1}$, where $\Lambda$ is a diagonal matrix with non-increasing eigen-values $1 = \lambda_0 \geq \lambda_1 \geq \cdots \geq \lambda_{N-1}$. Finally, we define the mapping $x_i \mapsto [\lambda_2^t \Phi[i,1] \quad \cdots \quad \lambda_d^t \Phi[i,d]]$, where $d$ is the desired dimensionality of the representation of the latent space, and $t$ is a hyperparameter which controls the duration of the diffusion process [26]. Note that we ignore the first eigenvector $\phi_0$ associated with $\lambda_0$, since it is a constant vector (due to the row normalization). In the latent space representation, the phase transition can be detected as sharp changes in the variation along the manifold. Such sharp changes occur at the phase transitions, which facilitate their detection with good resolution in the parameter domain.

We begin by demonstrating our method on simulated data described by the theoretical Haldane model, which is known to have a topological phase transition [2]. We show how our approach can distinguish between the two phases, based only on partial data from the system. To that end, our demonstration relies on the Haldane model, which is an archetypical model of topological systems.

The Haldane model is described by a honeycomb lattice with Nearest-Neighbors (NN) and Next-Nearest-Neighbors (NNN) couplings that includes a non-zero phase. This "Haldane phase" represents an applied (or synthetic) magnetic field which breaks time-reversal symmetry and opens a topological bandgap with edge-states. On the other hand, if the honeycomb potential is staggered, that is, the honeycomb potential comprises of two triangular lattices of different on-site energies, the resultant potential reaks the inversion symmetry, which may close the topological gap and open a trivial one, thus causing a topological phase transition. The system is described by the following Hamiltonian:

$$H = t_1 \sum_{\langle n,m \rangle} a_n^\dagger a_m + t_2 \sum_{\langle\langle n,m \rangle\rangle} e^{-n v_{nm} \varphi} a_n^\dagger a_m + M \sum_n b_n a_n^\dagger a_n \quad (1)$$

where $a_n$ $(a_n^\dagger)$ is the annihilation (creation) operator of cite $n$, $t_1$ is the NN coupling and $t_2$ is the NNN coupling of the honeycomb lattice, $M$ is the on-site staggered potential, $b_n$ equal $+1(-1)$ for the $A(B)$ (triangular) sublattice, the Haldane phase $\varphi$ in the NNN coupling. Here, the magnitude of $\varphi$ is constant while its sign varies according to: $v_{n,m} = \text{sign}(d_n \times d_m)_z$ where $d_n$ and $d_m$ are the vectors along the two bonds constituting the NNN couplings. In this model, detuning the parameter $M$ corresponds to changing the ratio between inversion-symmetry-breaking and time-reversal-breaking. Thus, for small $M$ $(0 < M < M_{critical})$ time-reversal symmetry breaking is dominant and causes a topological band-gap. However, when $M = M_{critical}$ the band-gap closes completely, and above that value, the bandgap opens again but this time as a trivial bandgap [27].

To highlight the strength of our methodology, we demonstrate its ability to identify topological phase transitions based on data on bulk states only, without any data on edge states whatsoever. To study the phase transition with data from the bulk only, we simulate the propagation of an initial wavepacket $\psi_0$ in a finite lattice described by Eq.1. The initial wavepacket $\psi_i(0)$ is some random superposition of bulk-states, so that it does not include any edge states (whose propagation has unique features when the phase is topological), making the detection of the topological phase of the system harder (Fig.1b-e). Moreover, our initial wavepacket is essentially noise, which means that it looks like noise also at the output – without any noticeable distinct features from which one could identify the topological state of the system. The simulated propagation is for a time period $T = 10$, for different values of $M$ ranged uniformly between $M_1 = 0$ and $M_{100} = 10$, while $\varphi$ is fixed at $\pi/2$. Changing the value of $M_i$ can close the topological bandgap and open a trivial bandgap (and vice versa). We denote the discrete-time evolution of the initial wavepacket $\psi_i(0)$ with $X_i \in \mathbb{C}^{D \times N_T}$, where $D = 170$ is the number of sites, and $N_T = 1,000$ is the number of time-steps. Finally, we apply Diffusion Maps to the set $\{X_i\}_{i=1}^{100}$, with $d = 2$. The number of dimensions $d$ is chosen such that it is the minimal value that provides enough information for detection. Figure 1f shows the result displayed by a one-dimensional curve since effectively, there is only one degree of freedom, which is the value of $M$. The colors of the curve in Fig. 1f correspond to the different values, $M_i$. The single most important feature in this figure is the cusp, which marks the presence of a topological phase transition. The Diffusion Maps methodology identifies this phase transition from bulk data only, and the position of the cusp is exactly at the theoretical phase transition of the system. This result shows that the algorithm can detect the phase-transition even without the most significant feature of the topological state – the edge states. Moreover, when the observed data set is relatively small, a similar result can be obtained with a variant of the analysis that relies on the spectral flow of the kernels of diffusion maps [see Supplementary Material]

The result presented in Fig. 1 leads to the following question: how partial can the data be for the algorithm to detect the phase transition? To this end, we study the performance of the algorithm on data taken from a single eigenstate of the system, which represents the smallest possible part of the system's spectrum (Fig.2a). We repeat the same calculations as before, only this time the initial wavevector $\psi_0$ is an arbitrary single eigenstate of the Hamiltonian in Eq.1. Fig.2b shows the obtained representation using Diffusion Maps. As before, we can notice a cusp exactly at the theoretical phase transition. The appearance of a cusp at the phase transition facilitates automatic detection of the topological phase transition, which is offered by the Diffusion Maps methodology (Fig. 2b). Other methods that were studied numerically in the past can also offer a way to identify the phase transition, although using them on the Haldane model yields a gradual curve which (as we show later) does not pinpoint the phase transition [see Supplementary Material]. The other methods aim to provide a low-dimensional representation which preserve the pairwise distances between the high-dimnesional data. These approaches seem to lack the sensitivity required to detect the dynmical change in the high-dimensional data when the phase transition occurs. On the other hand, Difusion Maps captures global relationships in the data which enable him to identify the different dynamics before and after the phase transition. Up to this point, we have shown, on simulated data, that Diffusion Maps can identify topological phases even when the available data is extremely scarce, as is usually the case in experiments.

As mentioned earlier, one important application of machine learning methods is the ability to study the phase transitions in true experimental systems without the ability to train a network (no training set is available) and without a known model (prescribing the relation between "input" and "output") to rely on. In addition, true experimental data inevitably implies observation noise, which cannot be fully simulated – because experiments are never ideal, and many times noise is system-specific at least in part. For these reasons, it is extremely important to demonstrate the Diffusion Maps methodology on true experimental data and understand if it can work reliably or does it work only on ideal data that can only be taken from simulations. To this end, we demonstrate that our approach successfully detects phase transitions in an experimental optical system that undergoes a topological phase transition [28]. We show that our methodology can detect the phase transition not only in cases where the data includes edge-states (whose evolution is unique in topological systems, making the identification of the underlying topology much easier), but also when only bulk states are excited. The latter case is relevant for many experimental systems, in which the edge-states cannot be directly excited and shows the robustness of our approach. The experimental results describing the phase transition were published in [28], and we use raw experimental data provided to us by the authors of [28]. The optical system is composed of a lattice of evanescently coupled waveguides each with a single mode in the optical frequency range. By judiciously designing the 3D waveguide structure, it is possible to make the light evolve in the system as electrons do in the anomalous Floquet topological insulator [28,29]. The waveguides are made helical in the evolution (propagation) z and are arranged in a diagonal square lattice. In each unit cell, the helicity is shifted by $\pi/2$, giving rise to a topological phase transition in a line of a 2D parameter space which is spanned by the wavelength and lattice constant.

The measurements are carried out in two types of initial excitations. The first type is focused mostly on the edge of the system- exciting the edge-states directly and collecting the data on their evolution at the exit facet of the system (Fig.3b). The second type of initial excitation is focused on exciting a single site in the middle of the lattice – not exciting any edge states at all (Fig.3d). For each type of initial excitation, we image the light intensity at the output intensity of the lattice (Fig.3c,e), under different conditions of different lattice constants and different wavelength, forming a grid of measurements in parameter space (Fig.4a,c). We emphasize that we use intensity data only, without interferometric measurements that would have revealed the phase of the output wavepackets and could have made the algorithmic processing easier. We use the Diffusion Maps algorithm on the recorded output, that is, the intensity images both for the measurements containing edgestates (Fig.4b) and for those without edgestates and observe the produced low dimensional manifold (Fig.4d).

Technically, for each output image (Example in Fig.3c,f), we extract a vector $x_i \in \mathbb{R}^D$, where $D = 162$ is the number of sites, such that $x_i[n]$ contains the intensity of the $n$th site. We apply Diffusion Maps to $\{x_i\}_i$. Figure 4b displays the obtained Diffusion Maps representation with $d = 3$ when the excitation is on the edge. Similarly, the result for the excitation on the bulk is displayed in Fig. 4d. As in the simulated results, in both cases, we notice a cusp point in the obtained representation. This cusp is located very close to the theoretical phase transition (Fig.4a,c). We note that the resulting curve in the simulated model and experimental model are quite different due to the high level of disorder in the experimental data, as often happens in many experiments. The Diffusion Maps methodology is robust enough to overcome these challenges, as demonstrated in

Fig. 4b,d. Diffusion Maps can detect a topological phase transition when the edge-states are not excited, even from dilute data, and under true experimental conditions that inevitably contain disorder and all kinds of experimental errors on the data. Figure 4d also displays an additional obtuse cusp, which is not as pronounced as the cusp indicating the topological phase transition associated bwith the Haldane model. This additional (obtuse) cusp is related to the destruction of the topoligcal phase due to enhanced NNN coupling near the edge of the paremeter range explored in the experimental system. Interestingly, Diffusion maps was able to point to this unexpected behavior (which is not covered by the physical model), with an obtuse cusp.

To summarize, Diffusion Maps seems to be a promising tool for detecting topological phase transitions both from simulated and experimental data. They can provide a powerful analysis tool for systems with limited access to the entire system and even without any data on the most distinct topological features – the edge states. Likewise, we have shown that Diffusion Maps can overcome the errors that are inevitable in any experiment, and still identify the topological phase transition based on limited data. Our experiment was carried out in the context of photonics, but the approach is general and could work for any experimental system. Although we study a certain class of topological phase transition, it is reasonable to conjecture that this approach can be extended to other types of phase transitions – both topological and non-topological. Finally, the big challenge is to employ Diffusion Maps to identify phase transitions in quantum many-body systems (such as many-body localization [30–32]), where theory can give only a limited indication of what to expect and what to examine in the experimental data. Indeed, the hope is that machine learning will offer an avenue to discoveries based on analyzing experimental data from quantum many-body systems.

Figure 1

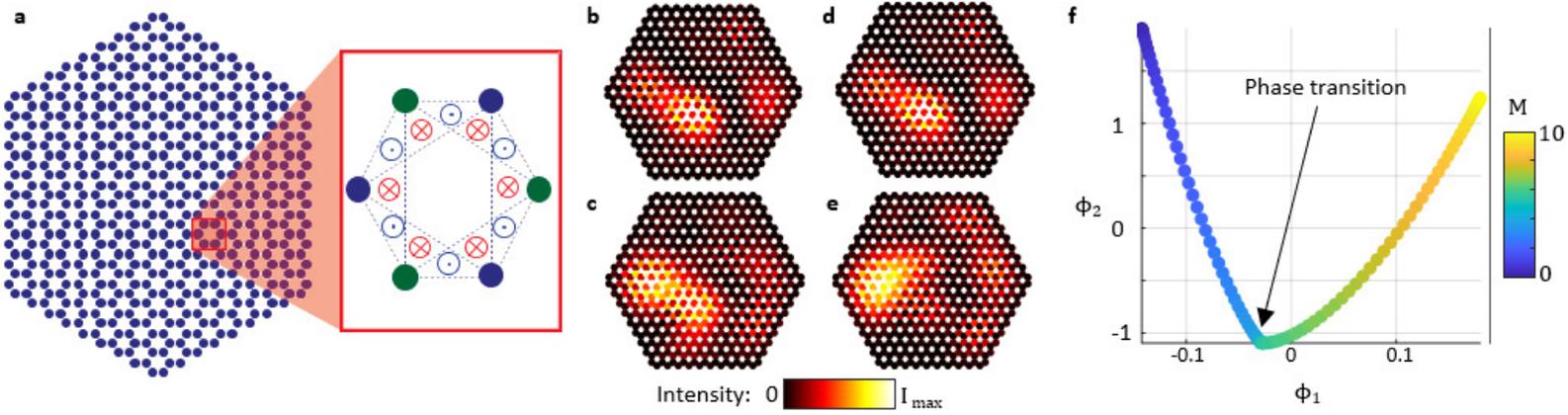

Figure 1. **(a)** The Haldane unit cell: the green and blue lattice sites indicate the two underlying sublattices whose potential may be detuned from one another. The red and blue inward and outward arrows are the non-uniform magnetic field. **(b,d)** Examples of the evolution of an arbitrary initial wavepacket made of bulk states only, within the Haldane model in the topological and the trivial cases, respectively (low and high detuning). The wavepacket is highlighted in green (brightness indicates the local aplitude) overlayed on the lattice (blue). **(c,e)** The states initiated in (b,d) after time $T$, respectively. **(f)** Scatter plot of the first two principal components of Diffusion Maps in the first simulation. The points are colored by the detuning in the honeycomb potential (magnitude of the detuning given by the color map). The cusp in **(f)**, found by applying Diffusion Maps to strictly bulk data (as (c,e)), is at the phase transition predicted by theory.

Figure 2

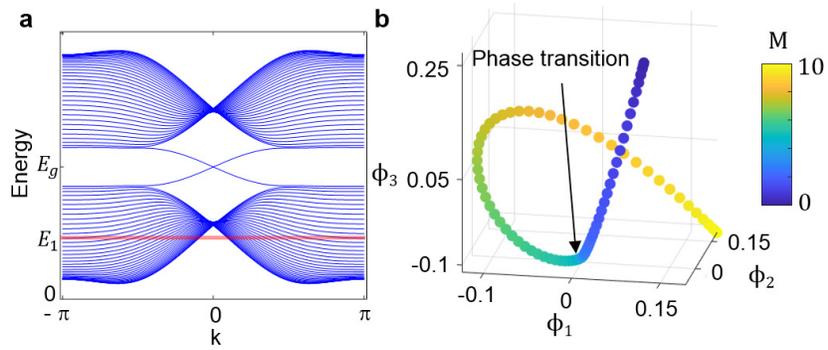

Figure 2. (**a**) The spectrum of the Haldane model. The available data is associated with a small part of the spectrum around $E_1$, which is represented by the red shaded area. (**b**) Diffusion Maps for the Haldane model, when observing the evolution of a single band eigenstate. The phase transition is at the sharp turn of the curve.

Figure 3

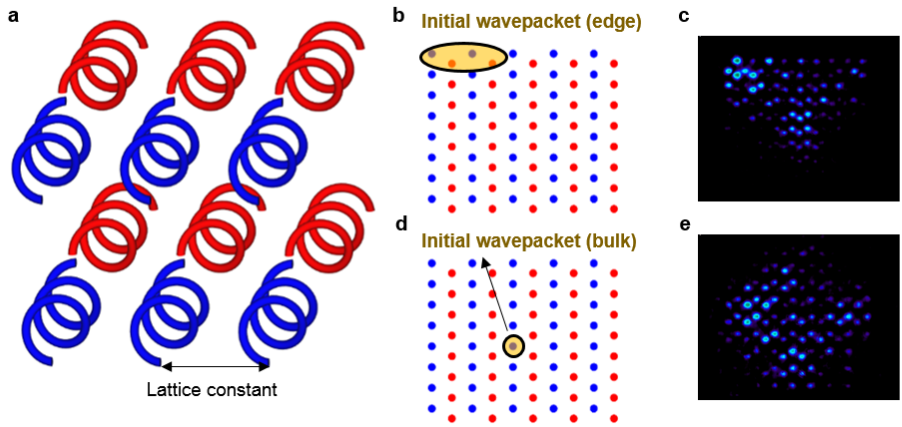

Figure 3: **(a)** Helical waveguide array of the detuned honeycomb lattice (as in [13]). **(b,e)** location of input beam when the edgestate are excited (b) and when the edgestates are not excited (c). **(c,e)** example of experimental output intensity images, when the input wavepacket is on the upper edge (c), and when the input is in the middle of the lattice (f).

Figure 4

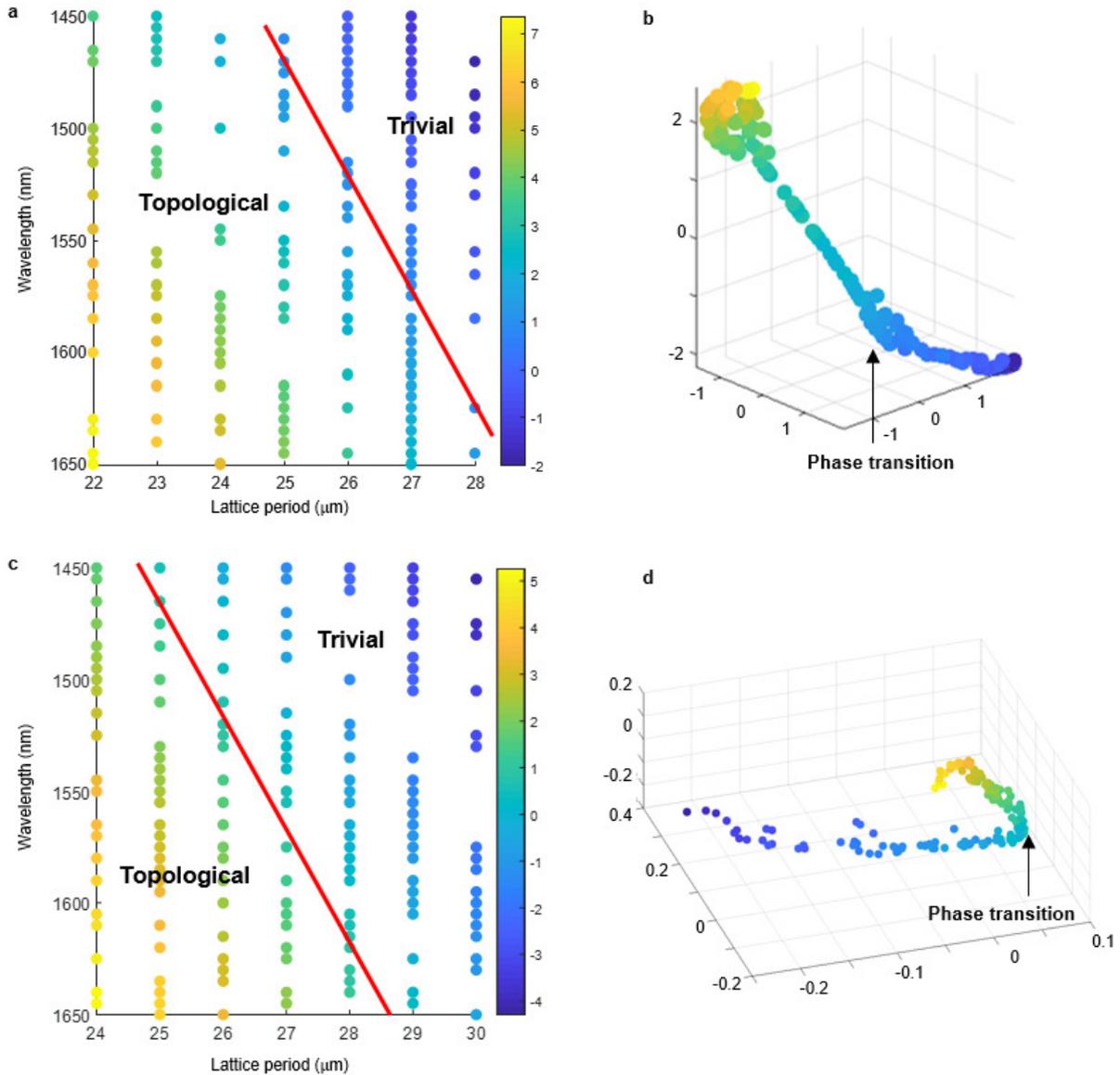

Figure 4. **(a)** Data distribution in parameter space for the case of edge excitation (Fig.3b-c); each point represents an experimental image. The color–bar represents the distance from the phase transition that is marked with a red line. **(b)** The first three principal directions obtained using Diffusion Maps applied to the experimental data. The data points in colors are the same as in (a). **(c-d)** same as (a-b) only for the case where the bulk is excited (Fig.3b-c).